\documentclass{article}

\usepackage[latin1]{inputenc}
\usepackage{graphicx}
\usepackage{amssymb,amsfonts,amsmath}
\usepackage{html}

\begin{document}

\title{Invariant manifolds of the Bonhoeffer-van der Pol oscillator}
\author{R. Ben\'{\i}tez$^1$, V. J. Bol\'os$^2$ \\
%EndAName
\\
{\small $^1$ Departamento de Matem\'aticas,}\\
{\small Centro Universitario de Plasencia, Universidad Extremadura.}\\
{\small Avda. Virgen del Puerto 2, 10600 Plasencia, Spain.}\\
{\small e-mail\textup{: \texttt{rbenitez@unex.es}}} \\
\\
{\small $^2$ Departamento de Matem\'aticas,}\\
{\small Facultad de Ciencias, Universidad Extremadura.}\\
{\small Avda. de Elvas s/n, 06071 Badajoz, Spain.}\\
{\small e-mail\textup{: \texttt{vjbolos@unex.es}}}}
\date{October 2007}

\maketitle

\begin{abstract}  The stable and unstable manifolds of a saddle fixed point (SFP) of
  the Bonhoeffer-van der Pol oscillator are numerically studied. A
  correspondence between the existence of homoclinic tangencies (which
  are related to the creation or destruction of Smale horseshoes) and
  the chaos observed in the bifurcation diagram is described. It is
  observed that in the non-chaotic zones of the bifurcation diagram,
  there may or may not be Smale horseshoes, but there are no
  homoclinic tangencies.

\end{abstract}


\section{Introduction}
The Bonhoeffer van der Pol oscillator (BvP) is the non-autonomous
planar system
\begin{equation}
  \label{eq:bvdp}
\left. \begin{array}{rcl}
  x'&=&x-\displaystyle{\frac{x^3}{3}}-y+I(t) \\
  y'&=&c(x+a-by)\\
\end{array} \right\} ,
\end{equation}
being  $a$, $b$, $c$ real parameters, and $I\left( t\right) $ an
external forcement. We shall consider only a periodic forcement
$I(t)=A\cos \left( 2\pi t\right) $ and the specific values for the
parameters $a=0.7$, $b=0.8$, $c=0.1$. These values were considered
in \cite{Raja96} because of their physical and biological
importance (see \cite{Scot77}).

In previous works \cite{Raja92}, the existence of ``horseshoe
chaos'' in BvP was studied analitically by means of the Melnikov
method applied to an equivalent system, the Duffing-van der Pol
oscillator (DvP). It was concluded that the BvP system should have
Smale horseshoes.  Nevertheless, the method used there can be
applied only for $b>1$ and $c<1/b$; which is not our case. In this
work we will show the relation between the chaos transitions in
the BvP system (\ref{eq:bvdp}) and the creation or destruction of
Smale horseshoes. Such horseshoes will be identified by the
existence of homoclinic tangencies between the invariant manifolds
of a saddle fixed point of the Poincaré map.  With this aim we
perform a numerical descriptive analysis of the stable and
unstable manifolds.

% The Appendices part is started with the command \appendix;
% appendix sections are then done as normal sections
% \appendix

\section{The bifurcation diagram}
\label{sec:bif} 
\begin{figure}
    \centering
    {\includegraphics[width=0.8\textwidth]{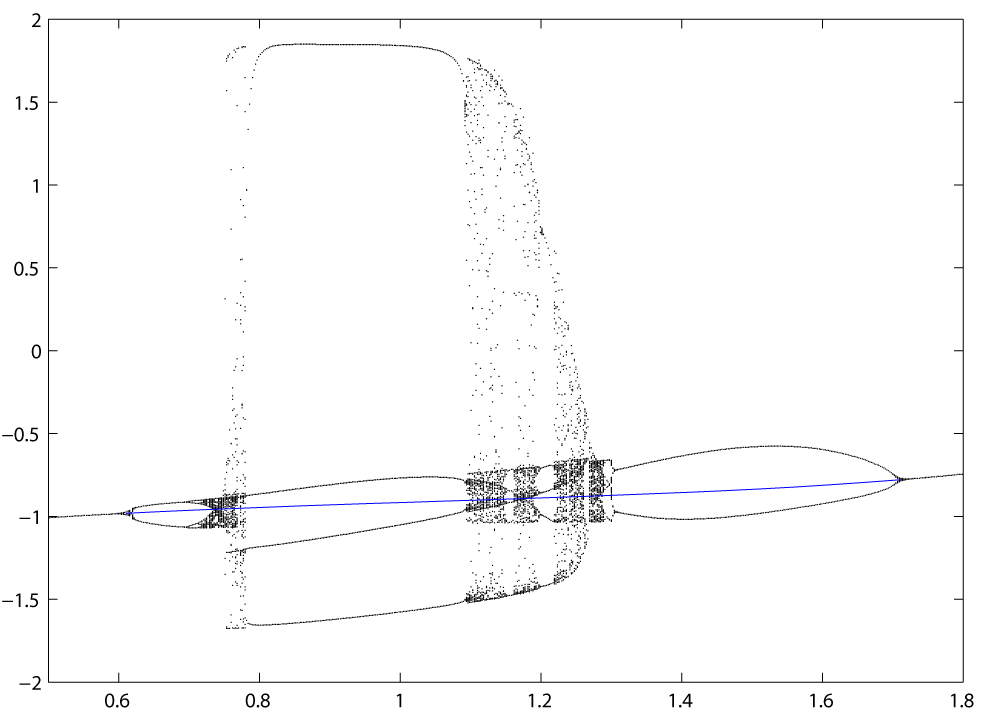}}
    \caption{Bifurcation diagram for the first coordinate of the
      periodic points of the Poincaré map, and $0.5\leq A\leq 1.8$.
      The blue line represents the position of the Saddle Fixed
      Point (SFP).}
          \label{fig:1}
\end{figure}

Obtaining, for different values of the parameter
$A$, the periodic fixed points of the Poincaré map (which is
defined by the flow of the system  on $t=1$, see \cite{Guck86}), a
typical bifurcation diagram is found, with chaotic and non chaotic
zones (see Fig. \ref{fig:1}). Such diagram has been deeply studied
in \cite{Wang89} and \cite{Raja96}.

\begin{figure}
  \centering
  {\includegraphics[width=0.9\textwidth]{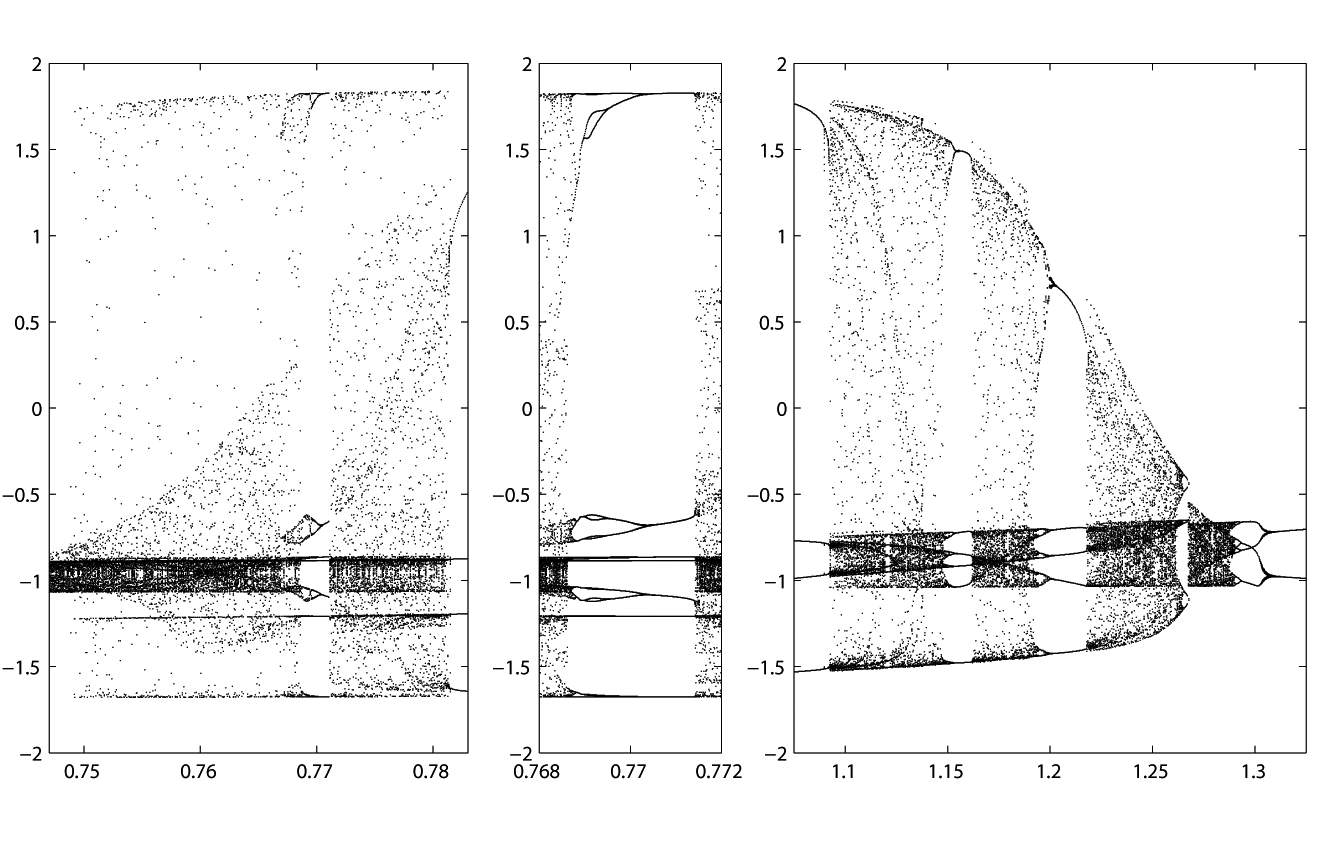}}
  \caption{Some details of the bifurcation diagram of the
    $x$-coordinate.}\label{bifz}
\end{figure}

For $0\lessapprox A\lessapprox 0.61$ and for $1.72 \lessapprox A$,
there is a unique attracting fixed point. The first bifurcation takes
place in $A\approx 0.61$, where two attracting 2-periodic points
appear, and a saddle fixed point (SFP) between them. Our aim is to
study the invariant manifolds of such SFP.

For $0.61\lessapprox A\lessapprox 0.735$ the bifurcation diagram
is very simple, from a dynamical point of view. The attracting
periodic points keep doubling its periods and forming a typical
non-chaotic bifurcation diagram. Their attraction basins are
perfectly separated and no fractal structures are formed.

For $0.735\lessapprox A\lessapprox 1.2835$ there is a sequence of
chaotic and non-chaotic zones (see Fig. \ref{bifz}). The largest
non-chaotic zone yields for $0.782\lessapprox A\lessapprox 1.092$,
and there is a 4-periodic attracting point. For example, the four
periodic points in the case $A=0.85$ are $x_1\approx
(-1.6444,0.4723)$, $x_2\approx (-1.1557,-0.2346)$, $x_3\approx
(-0.8362,-0.4906)$, $x_4\approx (1.8481,0.4416)$, and the image of
any of them is the next one, i.e.  $f(x_{i\mod 4})=x_{i+1\mod4}$.

Another event that must be mentioned is the sudden expansion of the
attractor that takes place at $A\approx 0.748$; once the horseshoe
chaos has begun. Shuch expansion has been deeply studied in
\cite{Raja96} using dynamical structure functions.

For $1.2835\lessapprox A\lessapprox 1.72$ there is no horseshoe
chaos. In this zone there are attracting periodic points.  This
situation is held until $A\approx 1.72$, where the SFP vanishes
and a unique attracting fixed point appears.

\section{Invariant manifolds}\label{sec3}

As we have mentioned above, our aim is to relate the chaos
transitions in the bifurcation diagram to the creation and
destruction of Smale horseshoes, which we shall identify with the
existence of homoclinic tangencies between the invariant
manifolds. In this section we are going to set the basic
definitions and the notation related to the invariant manifods;
then we shall describe the structure of the invariant manifolds,
depending on the value of the parameter $A$.

Let $x_0$ be a SFP of a discrete dynamical system with a
two-dimensional state space, given by a continuous map
$f:\mathbb{R} ^2\rightarrow \mathbb{R} ^2$. The stable manifold of
$x_0$, $W^s\left(
  x_0\right) $, is defined as
\[
W^s\left( x_0\right) :=\left\{ x\in \mathbb{R} ^2\quad : \quad
\lim _{n\rightarrow \infty}f^n\left( x\right) =x_0\right\} .
\]

On the other hand, the unstable manifold of $x_0$, $W^u\left(
  x_0\right) $, is defined as
\[
W^u\left( x_0\right) :=\left\{ x\in \mathbb{R} ^2\quad :\quad \lim
_{n\rightarrow \infty }f^{-n}\left( x\right) =x_0 \right\} .
\]

In this case, these invariant manifolds are one-dimensional.
Moreover, it is obvious that if they intersect at one point
different from the SFP, then they must intersect at an infinite
set of points and Smale horseshoes are formed (see \cite{Guck86}).

\begin{figure}
  \centering
  {\includegraphics[width=10cm]{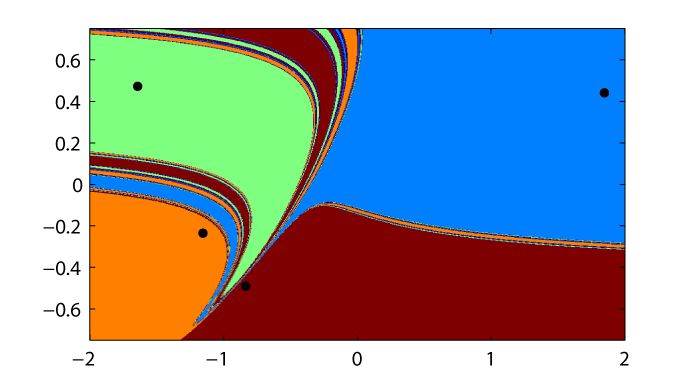}}
  \caption{Attraction basins of the four attracting periodic points
    for $A=0.85$. For such $A$, the SFP lies approximately in
    $(-0.9384,-0.4146)$, which is in the boundary of the four
    attraction basins.}\label{cuencas085}
\end{figure}

A simple consequence of these definitions is that, in presence of
several attracting fixed points, $W^s\left( x_0\right)$ is
included in the boundary of the attraction basins of these points.
Thus, if there are horseshoes, the attraction basins present a
fractal structure (see Fig. \ref{cuencas085}).

Each invariant manifold has two branches that ``arrive'' at the
SFP (in the case of the stable manifold) or ``leave'' the SFP (for
the unstable manifold). More precisely, let $x\in W^s\left(
x_0\right)$, then $d^s(f^n(x),x_0)\to 0$ as $n\to\infty$, being
$d^s$ the arclength distance defined on $ W^s\left( x_0\right)$.
On the other hand, if $x\in W^u\left( x_0\right)$, then
$d^u(f^n(x),x_0)\to \infty$, being $d^u$ the arclength distance
defined on $ W^u\left( x_0\right)$ (see \cite{Engl04}).

In our case, the vector field $f$ is given by the Poincaré map
defined by the flow of the system (\ref{eq:bvdp}) on $t=1$. This
map has a unique SFP which appears with the first bifurcation at
$A\approx 0.61$ and vanishes with the last bifurcation,
approximately at $A\approx 1.72$ (see Fig. \ref{fig:1}). Therefore
we are going to focus on the values $0.61\lessapprox A\lessapprox
1.72$, which are the ones for which the invariant manifolds exist.

\subsection{The stable manifold}\label{seclaquesea}

The simplest method for plotting the stable manifold is the well known
``inverse method'', which consists in plotting the unstable manifold
of the inverse of the Poincaré map.

With this method, taking into account the machine precision we are
using, we obtain that the branches of the stable manifold reach a
numerical infinity at a finite time. This ``out of range'' takes place
only for the $x$ coordinate, due to the cubic term of (\ref{eq:bvdp}).
Nevertheless it is important to remark that, in theory, no point on
the stable manifold can go to infinity at a finite time.

\begin{figure}
  \centering
  {\includegraphics[width=0.9\textwidth]{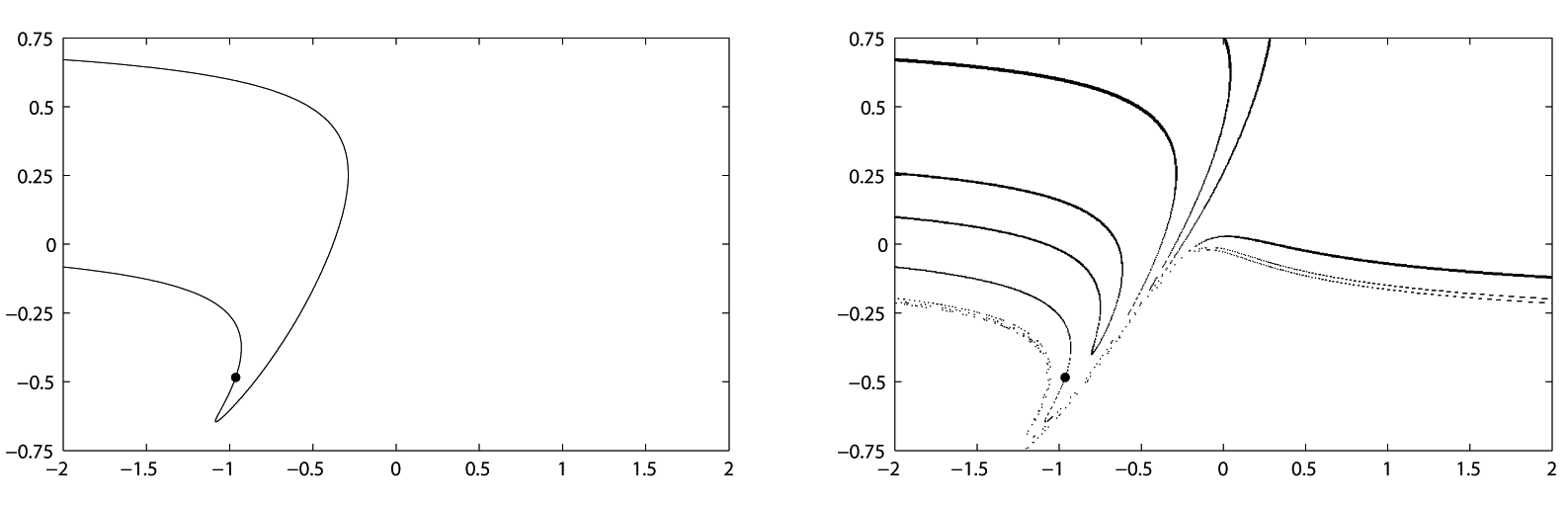}}
  \caption{Stable manifold for $A=0.70$. Left: branches of the
    manifold plotted with the inverse method, before they get out of
    range.  Right: stable manifold plotted with the method of
    ``convergent points''. The SFP is marked with a
    black circle.}\label{est070}
\end{figure}

Because of this behaviour, it is not possible to draw numerically
the whole manifold using the inverse method. In fact, it only
works until the branches are out of range (Fig. \ref{est070}
left).

Recently, new algorithms for plotting stable manifolds without
computing the inverse have been developed (see
\cite{Engl04,Hobs91,Kost96}). However this methods cannot be used
here because they require that the manifolds are sufficiently
bounded.

To avoid these problems, in order to plot the stable manifold, we
use another complementary method which consists in finding the
points that, after a given number of iterations, lie near the SFP
(Fig. \ref{est070} right).  Nevertheless this method is
computationally more expensive. For more details refer to Section
\ref{secalg}.

\subsection{The unstable manifold}

The unstable manifold remains in a bounded region close to the SFP.
However, the branches of the manifold present some folds that, at
first sight, look like vertices, difficulting its study.

\begin{figure}
  \centering
  {\includegraphics[width=0.6\textwidth]{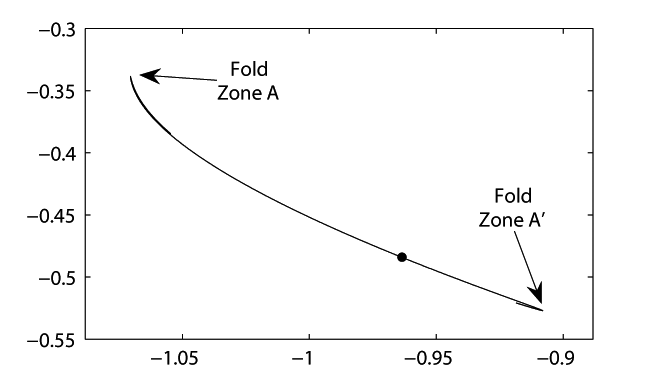}}
  \caption{Unstable manifold for $A=0.70$. The SFP is marked with a
    black circle.}\label{in070}
\end{figure}

\begin{figure}
  \centering {\includegraphics[width=\textwidth]{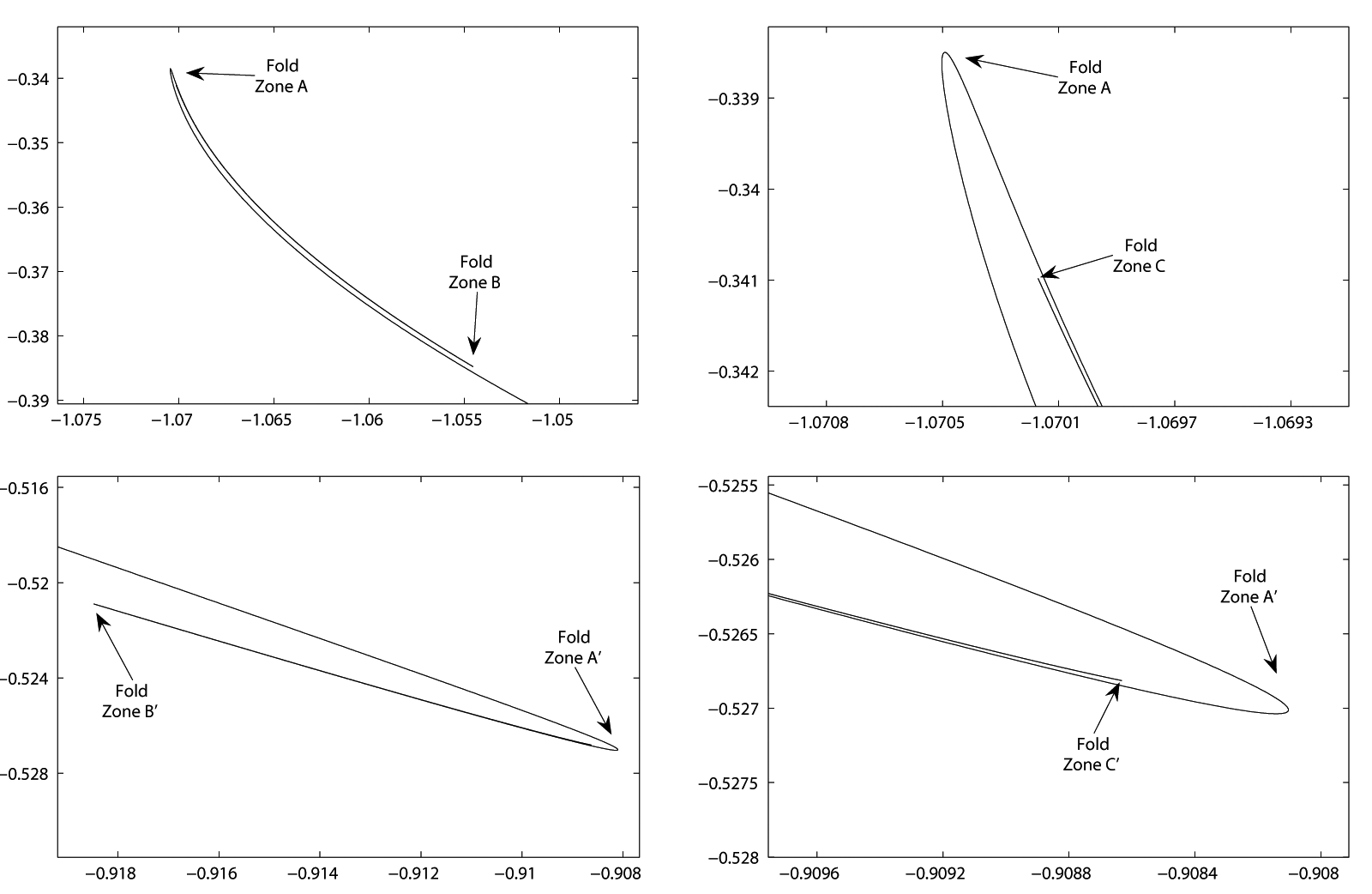}}
  \caption{Details of the unstable manifold for $A=0.70$.}\label{in070z}
\end{figure}

\begin{figure}
  \centering {\includegraphics[width=\textwidth]{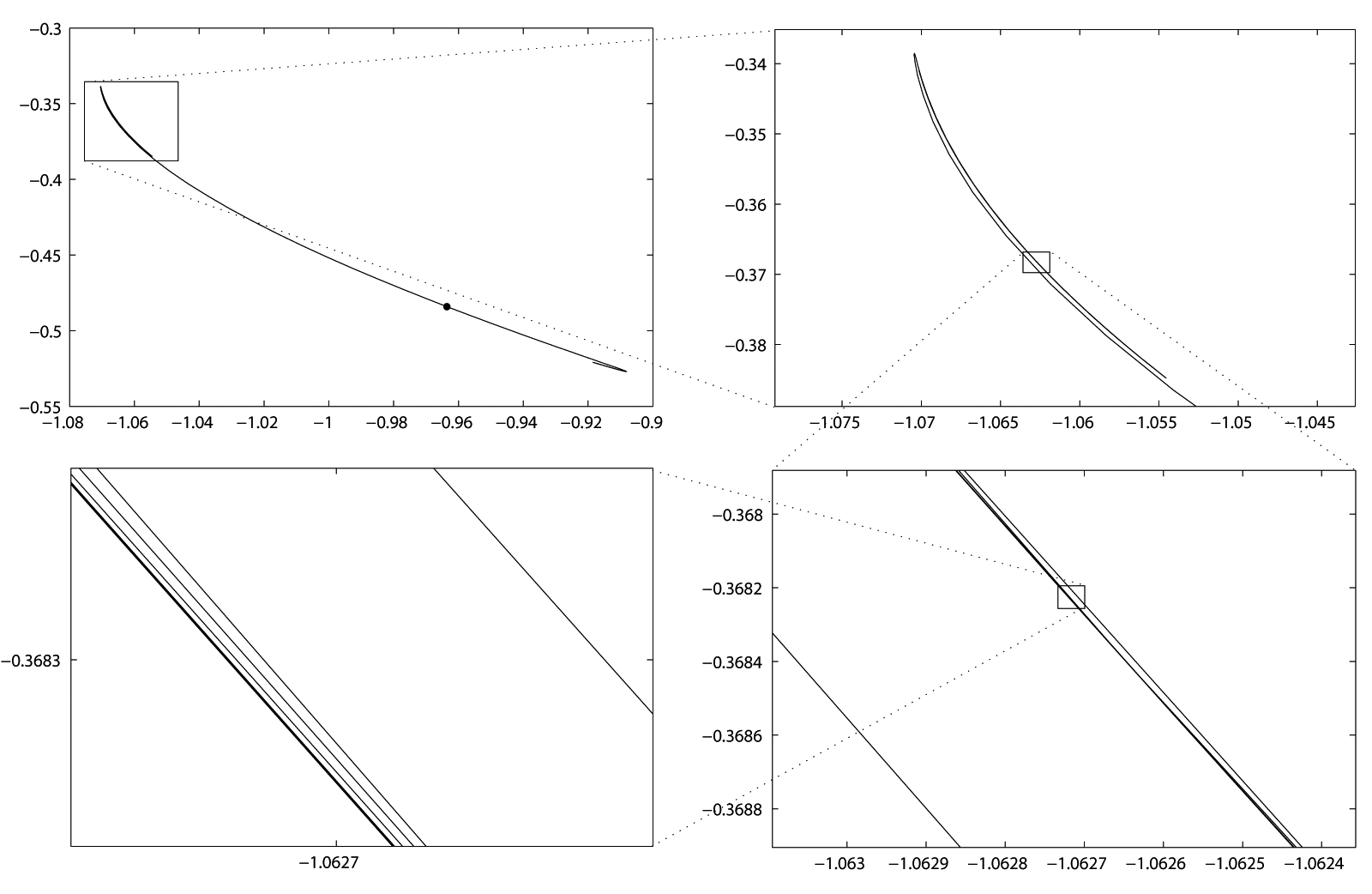}}
  \caption{Details of the unstable manifold for $A=0.70$.}\label{in070z2}
\end{figure}

For example, for $A=0.70$, the SFP is at $(-0.9635,-0.4841)$,
approximately.  One branch arrives to the zone near
$(-1.0704,-0.3385)$ where it folds (fold zone A). Other fold zones
that have to be mentioned are near the points $(-1.0545, -0.3848)$
(fold zone B) and $(-1.0701,-0.341)$ (fold zone C). Since our
Poincaré map is orientation reversing, the other branch of the
manifold has an analogous structure, presenting the fold zones A',
B', and C', as the images of the fold zones A, B, and C. The
branches are ``trapped'' between the fold zones B-C and B'-C'
respectively (see Figures \ref{in070}, \ref{in070z} and
\ref{in070z2}).

\begin{figure}
  \centering
  {\includegraphics[width=.6\textwidth]{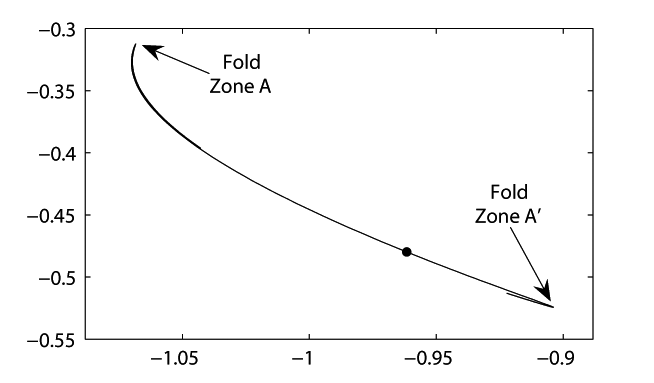}}
  \caption{Unstable manifold for $A=0.71$. The SFP is marked with a
    black circle.}\label{in071}
\end{figure}

\begin{figure}
  \centering
  {\includegraphics[width=\textwidth]{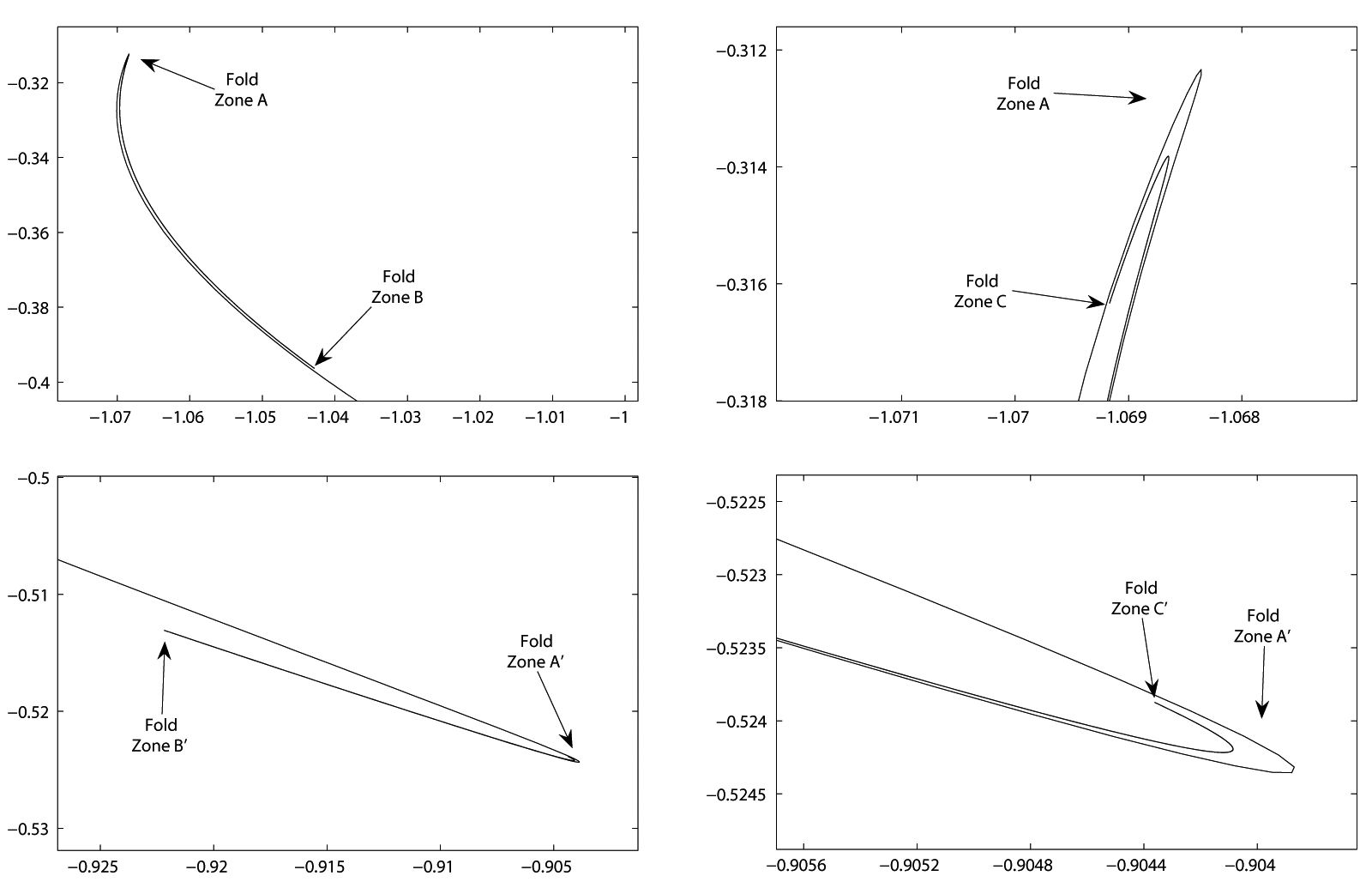}}
  \caption{Details of the unstable manifold for
    $A=0.71$.}\label{in071z}
\end{figure}

\begin{figure}
  \centering {\includegraphics[width=\textwidth]{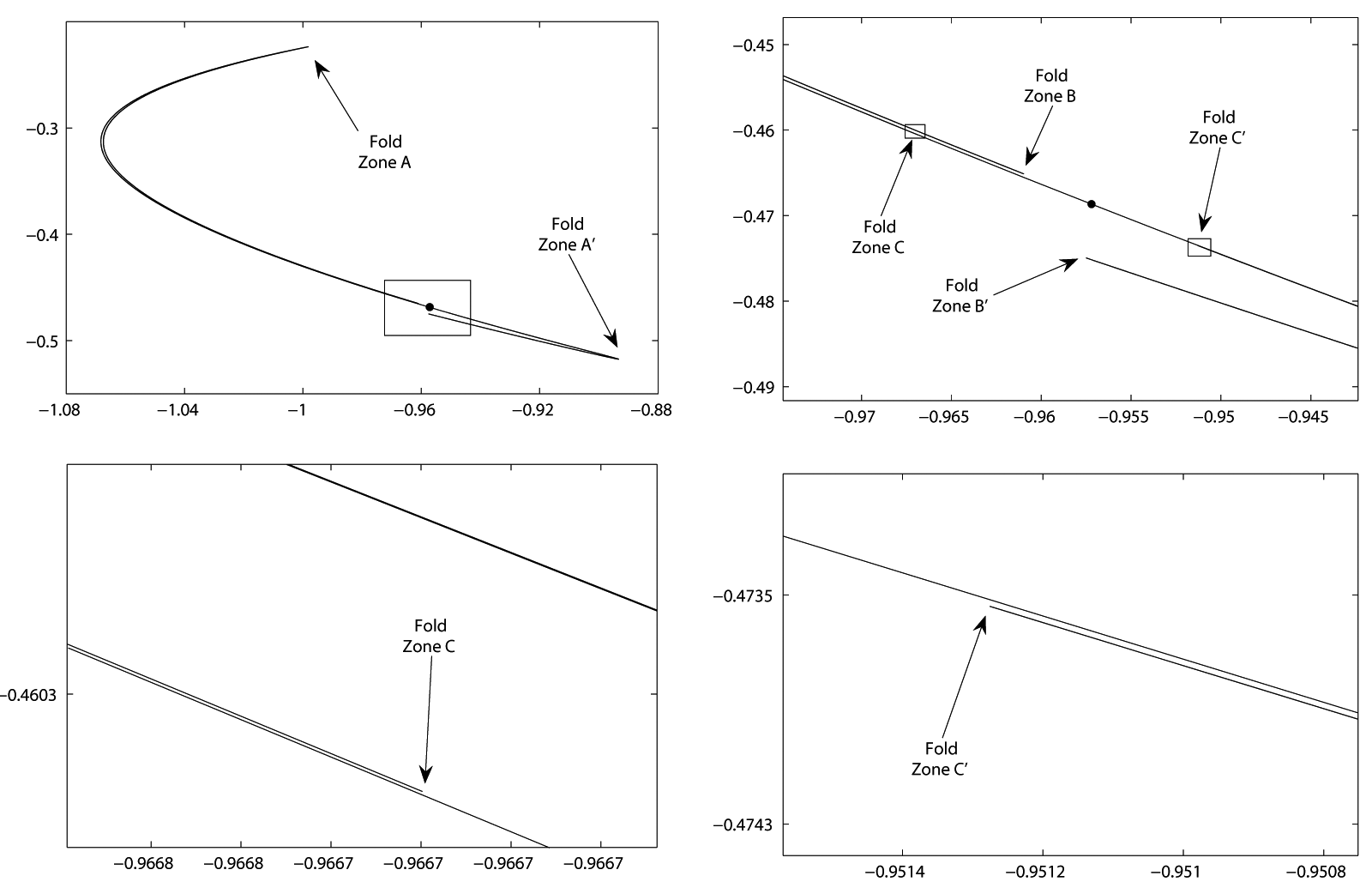}}
  \caption{Unstable manifold for $A=0.735$. The SFP is marked with a
    black circle.}\label{in0735}
\end{figure}

\begin{figure}
  \centering
  {\includegraphics[width=\textwidth]{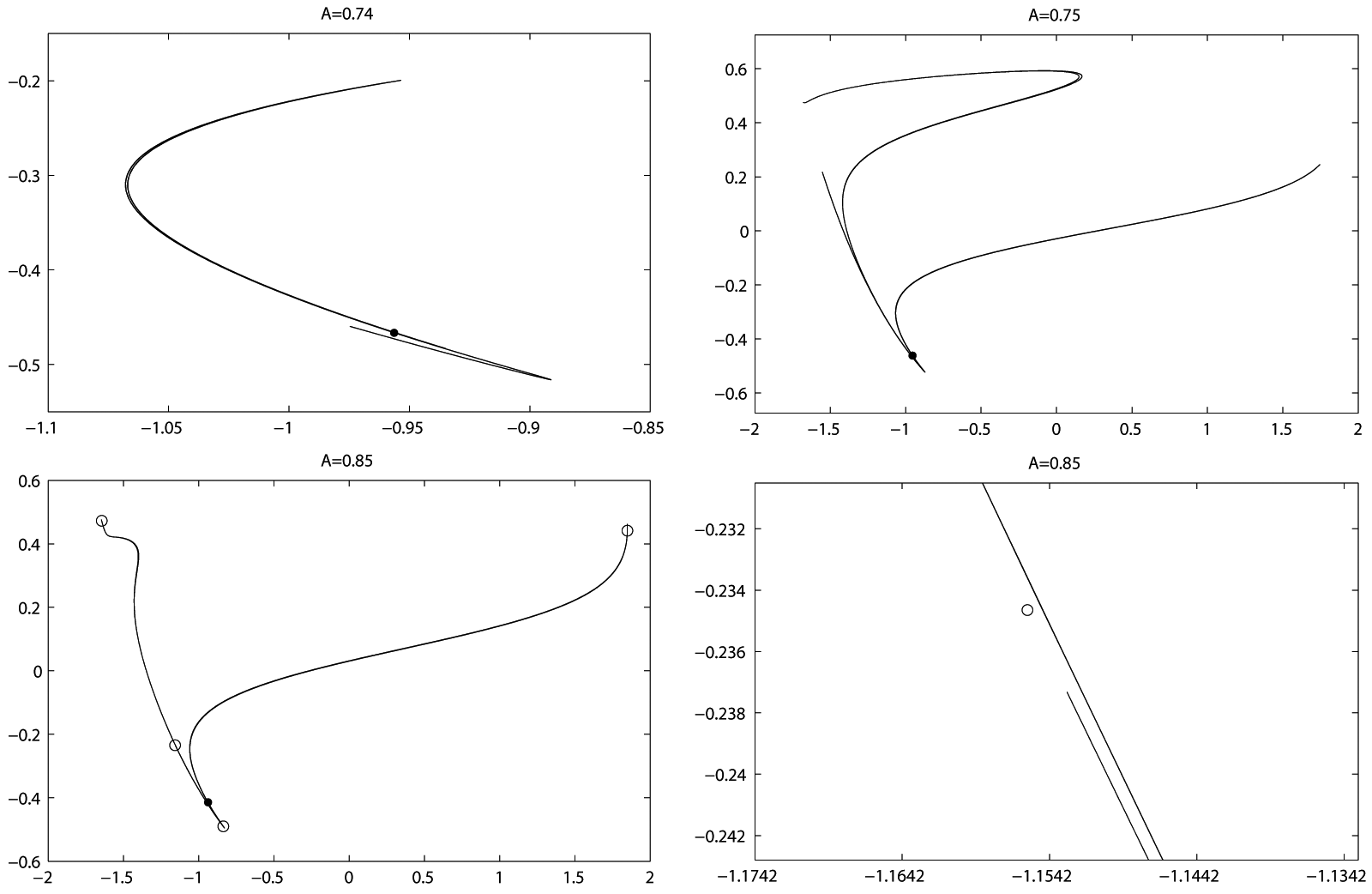}}
  \caption{Unstable manifolds for $A=0.74$, $A=0.75$, and $A=0.85$.
    The SFP is marked with a black circle. For $A=0.85$, the four
    attracting periodic points (marked with circles) are near the
    main fold zones.}\label{in074_075_085}
\end{figure}

If the parameter $A$ is increased, the fold zones B, C, B', C'
approach the SFP (Figures \ref{in071}, \ref{in071z}). When these
fold zones arrive near the SFP, the unstable and stable manifolds
will intersect, causing the first set of Smale horseshoes. For
example, if $A=0.735$, these four fold zones are very close to the
SFP (Fig. \ref{in0735}). For $A=0.74$ the first set of horseshoes
has been created (Fig. \ref{in074_075_085}).

As we have mentioned in Section \ref{sec:bif}, for $A\approx
0.748$ there is a sudden expansion of the size of the attractor,
and consequently of the size of the unstable manifold. For
example, this expansion is well described in Fig.
\ref{in074_075_085} for $A=0.75$. In this case the main fold zones
are near the points $(1.7458,0.2452)$, $(-1.6781,0.4752)$,
$(-1.2212,-0.2360)$, and $(-0.8721,-0.5232)$, which are the zones
of the bifurcation diagram with greater density of points.

In the non-chaotic big zone $0.782\lessapprox A\lessapprox 1.092$
there are four attracting points of period 4. These points are
near the main fold zones as we can see for $A=0.85$ in Fig.
\ref{in074_075_085}.

\section{Smale horseshoes}
\label{sec4}

\begin{figure}
  \centering
  {\includegraphics[width=\textwidth]{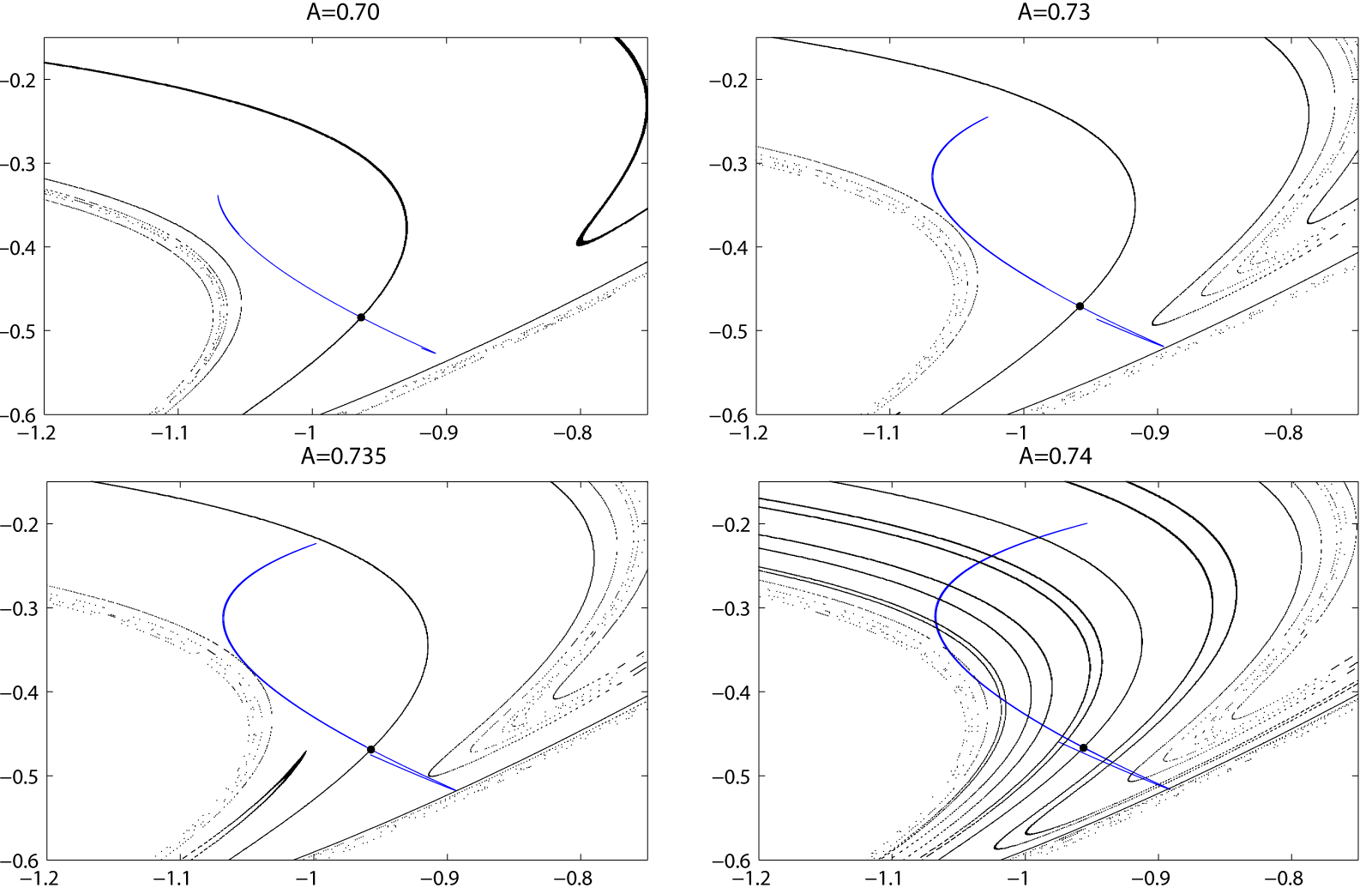}}
  \caption{Stable (black) and unstable (blue) manifolds for
    $A=0.70$, $A=0.73$, $A=0.735$, $A=0.74$. The SFP is marked with a
    black circle.}\label{estin070_073_0735_074}
\end{figure}

For $0.61\lessapprox A\lessapprox 0.735$ there are no
intersections between the stable and unstable manifolds. So there
is no horseshoe chaos in this region (Fig.
\ref{estin070_073_0735_074}).

For $A=0.735$ it is observed that several homoclinic tangencies are
about to be formed. This will cause the creation of the first set of
horseshoes and the first chaos transition. All the Smale horseshoes
are in the region $0.735\lessapprox A\lessapprox 1.2835$. These
horseshoes are ``quadruple'' since each branch of each invariant
manifold intersects the two branches of the other manifold.

\begin{figure}
  \centering
  {\includegraphics[width=\textwidth]{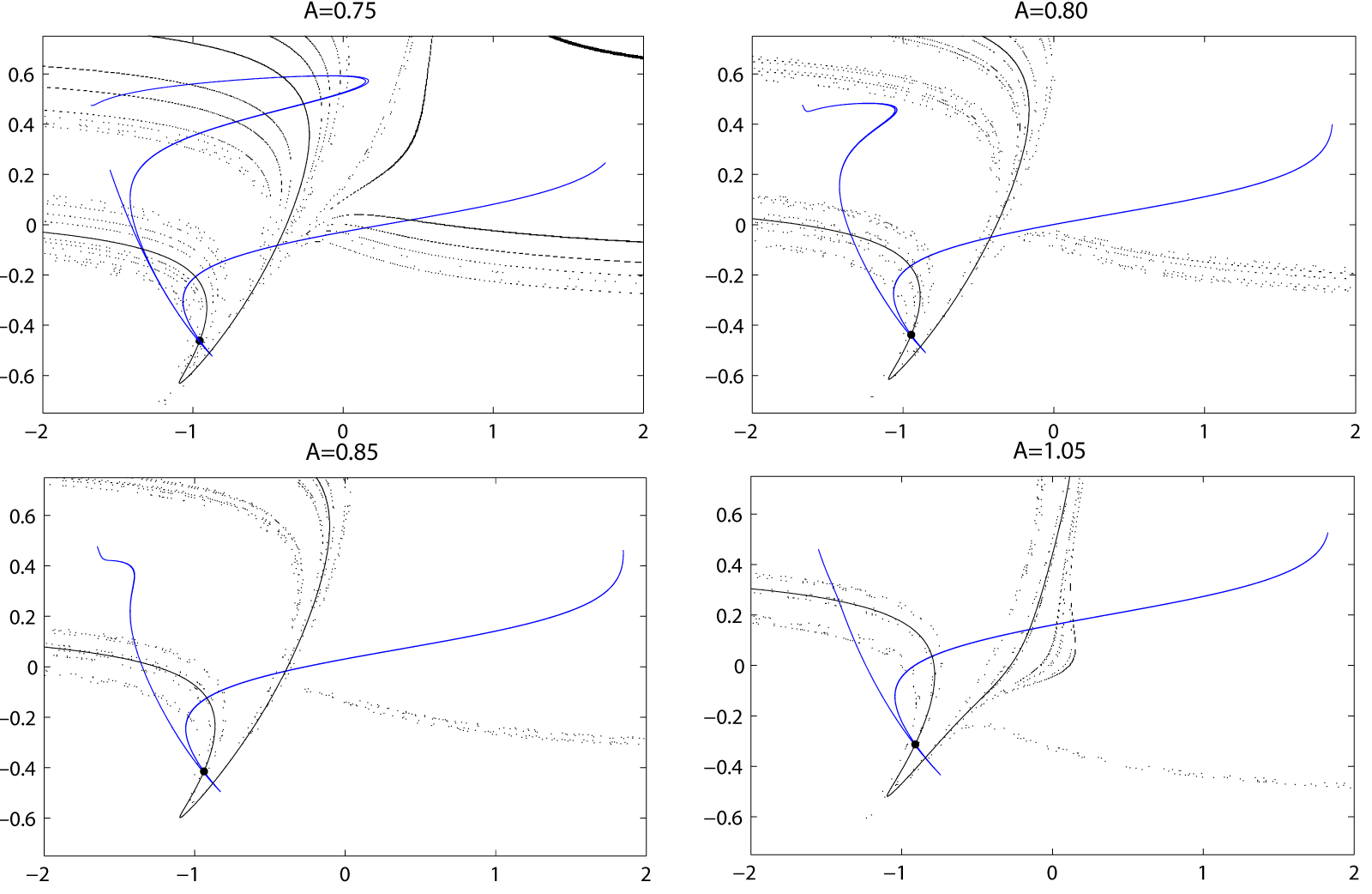}}
  \caption{Stable (black) and unstable (blue) manifolds for
    $A=0.75$, $A=0.80$, $A=0.85$, $A=1.05$. The SFP is marked with a
    black circle.}\label{estin075_080_085_105}
\end{figure}

\begin{figure}
  \centering
  {\includegraphics[width=\textwidth]{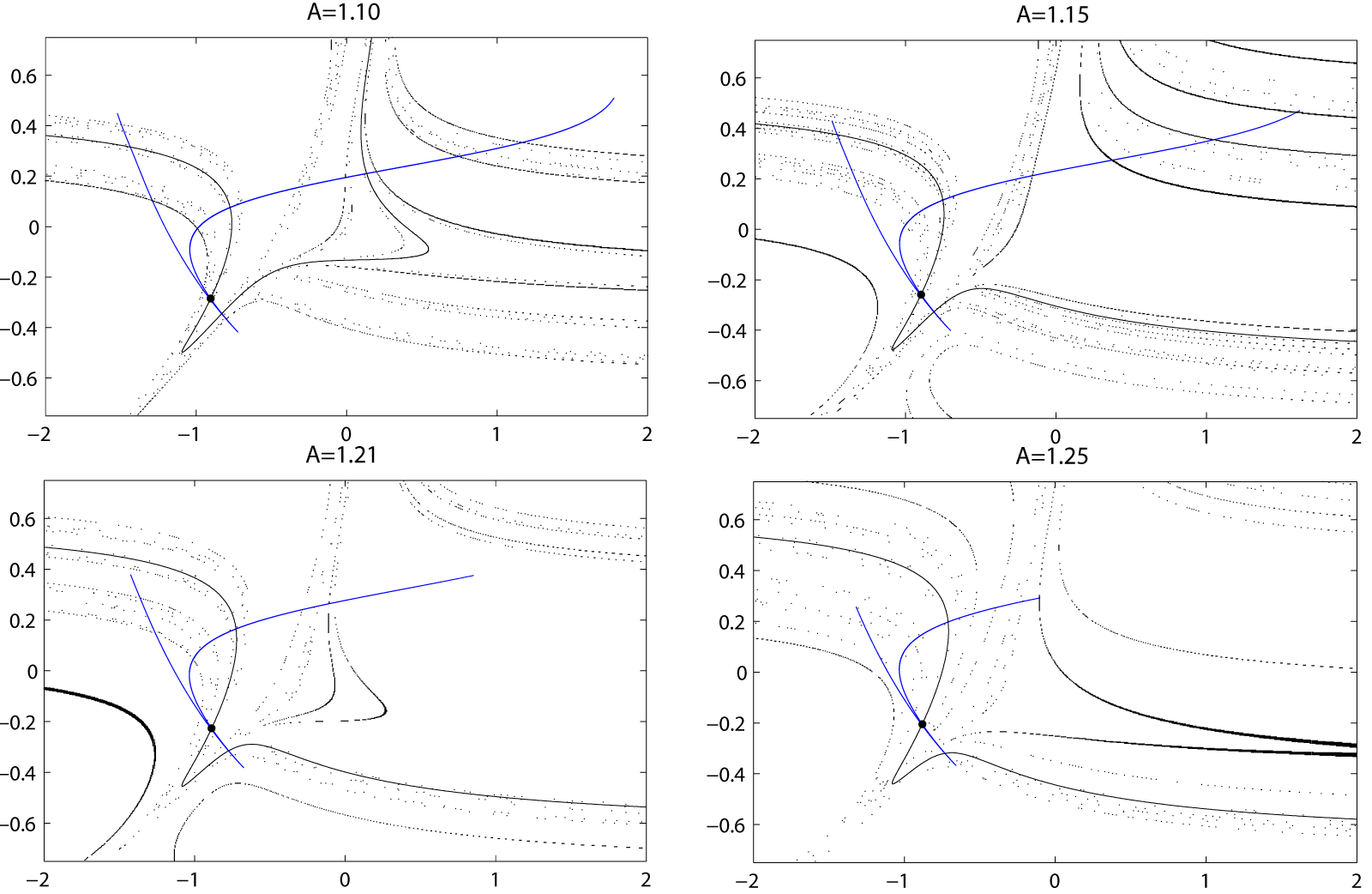}}
  \caption{Stable (black) and unstable (blue) manifolds for
    $A=1.10$, $A=1.15$, $A=1.21$, $A=1.25$. The SFP is marked with a
    black circle.}\label{estin110_115_121_125}
\end{figure}

For $A=0.74$ the first set of horseshoes has been already created
(Fig. \ref{estin070_073_0735_074}), and for $A\approx 0.748$
occurs the sudden expansion of the size of the attractor, causing
new intersections between the stable and unstable manifolds and
new horseshoes (Fig. \ref{estin075_080_085_105}, $A=0.75$). From
this point, chaotic and non-chaotic zones alternate, because
homoclinic tangencies between the stable and unstable manifolds
are not continuously formed. For example, for $A\approx 0.785$ the
chaos disappears because there are no homoclinic tangencies (Fig.
\ref{estin075_080_085_105}, $A=0.80$). This non-chaotic zone (Fig.
\ref{estin075_080_085_105}, $A=0.85$, $A=1.05$) remains until
$A\approx 1.11$ (Fig. \ref{estin110_115_121_125}, $A=1.10$,
$A=1.15$), when new homoclinic tangencies are formed. This
alternation of chaotic and nonchaotic zones is maintained until
$A\approx 1.2835$ (Fig. \ref{estin110_115_121_125}, $A=1.21$ for a
non-chaotic zone and $A=1.25$ for a chaotic zone).

\begin{figure}
  \centering
  {\includegraphics[width=\textwidth]{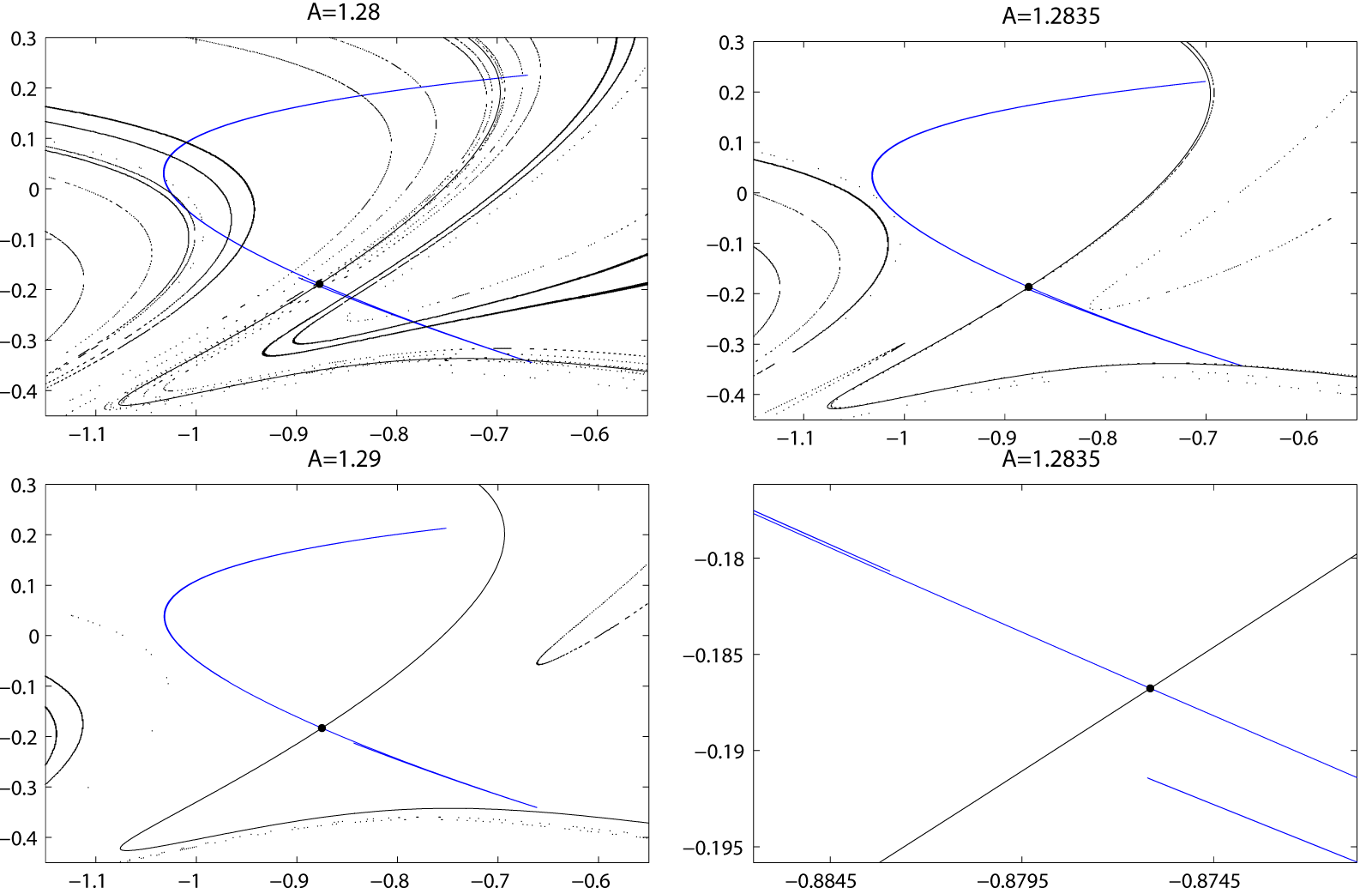}}
  \caption{Stable (black) and unstable (blue) manifolds for
    $A=1.28$, $A=1.2835$, $A=1.29$. The SFP is marked with a
    black circle.}\label{estin128_129}
\end{figure}

For $A\approx 1.2835$ the last set of horseshoes is undone (Fig.
\ref{estin128_129}), in a process inverse to the formation of
horseshoes for $A\approx 0.735$. In fact, for these two values of
$A$, the invariant manifolds are qualitatively similars.

Finally, for $1.2835\lessapprox A\lessapprox 1.72$ there is not
any horseshoe. Moreover, there are not homoclinic tangencies, and
therefore there is no chaos.

\section{Algorithms}\label{secalg}

\begin{itemize}
\item The Poincaré map, $f$, is computed solving the system
  (\ref{eq:bvdp}) using a Runge-Kutta method of order 4-5, with a
  local error of $10^{-12}$.
\item The SFP is obtained with an absolute error of $10^{-12}$, using
  a method, that we shall describe below, based on the fact that the
  iteration of a neighbourhood of the SFP is also a neighbourhood of
  the SFP.

  This algorithm is as follows: first consider an initial mesh $M_0$
  of a square interval $I_0=[x^1_0,x^2_0]\times [y^1_0,y^2_0]$ containing
  a unique SFP $(x^*,y^*)$.

  Then, given a positive $\epsilon_0$, we consider the set
  \[
  C_1 = \{(x,y)\in M_0: \|f(x,y)-(x^*,y^*)\|<\epsilon_0\}.
  \]

  Next we repeat this process with another mesh $M_1$ of the square
  interval $I_1=[x^1_1,x^2_1]\times [y^1_1,y^2_1]$, where
  \begin{eqnarray*}
  x^1_1&:=&\textrm{min}\left\{ x:\left( x,y\right) \in
  C_1\right\} ~~;~~
  x^2_1:=\textrm{max}\left\{ x:\left( x,y\right) \in
  C_1\right\} \\
  y^1_1&:=&\textrm{min}\left\{ y:\left( x,y\right) \in
  C_1\right\} ~~;~~
  y^2_1:=\textrm{max}\left\{ y:\left( x,y\right) \in
  C_1\right\} ,
  \end{eqnarray*}
  and for another positive $\epsilon_1<\epsilon_0$.  Obviously $I_1\subseteq
  I_0$.

  Thus, repeating this process, with a suitable choice of meshes $M_n$ and $\epsilon_n$,
  we can obtain a sequence of nested squared intervals
  $I_0\supset I_1\supset I_2\ldots$ such that
  \[
  (x^*,y^*)=\displaystyle{\bigcap_{n=0}^{+\infty}I_{n}}.
  \]

  For example, in our case, we have chosen the initial square interval
  $I_0=[-1.25,-0.25]\times [-0.6,0.4]$ and an initial mesh $M_0$ of
  $60\times 60$ points for every $A$, with $\epsilon _0=10^{-1}$.
  In the next iterations we have
  considered meshes of $30\times 30$ points, and $\epsilon_n=10^{-n-1}$.
  We have to note that, in this case, it is not necessary to use other
  methods based on model perturbations, as the control methods of OGY
  or Pyragas (see \cite{OGY90,Raja95,Rame01}).

\item In order to compute the invariant manifolds, first we estimate
  their tangential slopes at the SFP. This is done by using the
  eigenvectors of the Jacobian matrix $Df$ of the Poincaré map at the
  SFP.

\item To compute the unstable manifold we iterate a segment with
  length of order $10^{-3}$, centered at the SFP, with the appropriate slope
  previously found. The number of iterations may change with the value
  of $A$, but it is usually between 12 and 24.
\item For the stable manifold, two complementary methods have been used.
  \begin{itemize}
  \item First, we have iterated twice a small segment (with the
    appropiate slope), using the inverse of the Poincaré map.  As was
    pointed out in Section \ref{seclaquesea}, this ``inverse method''
    is only valid to compute the branches of the stable manifold
    until they are out of range. So the inital segment must be small
    enough to keep the iterations within range.
  \item Next, we find the points of a mesh such that, after $n$
    iterations, their distance to the SFP is less than a given
    $\epsilon$. In our case we used $n=8$, a mesh of $900\times 900$
    points, and $10^{-3}<\epsilon<10^{-2}$.

  \end{itemize}
\end{itemize}

\section{Final remarks}

The main results obtained are the following:

\begin{itemize}

\item The existence of Smale horseshoes in the forced BvP oscillator
  (\ref{eq:bvdp}) has been explicitly checked, depending on the
  amplitude of the external forcement $A$. In particular, the
  horseshoes appear for $A\in (0.735, 1.2835)$, approximately.
\item The chaotic zones of the bifurcation diagram are related to
  creation and/or destruction of horseshoes, i.e., to the existence of
  homoclinic tangencies between the invariant manifolds.
\item The sudden expansion of the attractor seems to be related to the
  creation of the first set of horseshoes.
\end{itemize}

\section*{Acknowledgments}
  We would like to thank Jos\'e A. Rodr\'{\i}guez, from the University of Oviedo for his valuable help. Both authors were partially supported by Junta de Extremadura, and MCYT-FEDER Grant Number MTM2004-06226.


\begin{thebibliography}{00}
\bibitem{Raja96} {\ S. Rajasekar.} Dynamical structure functions at
  critical bifurcations in a Bonhoeffer-van der Pol equation.
  \textit{Chaos, Solitons and Fractals} \textbf{7} (11) (1996),
  1799-1805.

\bibitem{Scot77} {\ A. C. Scott.} \textit{Neurophysics}. Wiley, New
  York (1977).

\bibitem{Raja92} {\ S. Rajasekar, S. Parthasarathy, M. Lakshmanan}
  Prediction of horseshoe chaos in BVP and DVP oscillators.
  \textit{Chaos, Solitons and Fractals} \textbf{2} (3) (1992),
  271-280.

\bibitem{Guck86} {\ J. Guckenheimer, P. Holmes}.  Nonlinear
  oscillations, dynamical systems, and bifurcations of vector fields.
\textit{Spinger-Verlag} New York (1986)

\bibitem{Wang89} {\ W. Wang.} Bifurcations and chaos of the
  Bonhoeffer-van der Pol model, \textit{J. Phys. A: Math. Gen.}
  \textbf{22} (1989), L627-L632.

\bibitem{Engl04} {\ J. P. England, B. Krauskopf, H. M. Osinga},
  Computing one-dimensional stable manifolds and stable sets of planar
  maps without the inverse, \textit{SIAM Journal of Applied Dynamical
    Systems} \textbf{3} (2) (2004), 161-190.

\bibitem{Hobs91} {\ D. Hobson}, An efficient method for computing
  invariant manifolds of planar maps, \textit{Journal of Computational
    Physics} \textbf{104} (1991), 14-22.

\bibitem{Kost96} {\ E. J. Kostelich, J. A. Yorke, Z. You}, Plotting
  stable manifolds: error estimates and noninvertible maps,
  \textit{Physica D} \textbf{92} (1996), 210-222.


\bibitem{OGY90} {\ E. Ott, C. Grebogi, J. Yorke}.
Controlling Chaos. \textit{Phys. Rev. Lett.} \textbf{66} (1990),
1196-1199.

\bibitem{Raja95} {\ S. Rajasekar}.
Controlling Unstable Periodic Orbits in a Bonhoeffer-van der Pol
equation. \textit{Chaos, Solitons and Fractals} \textbf{5} (1995),
2135-2144.

\bibitem{Rame01} {\ M. Ramesh, S. Narayanan}.
Chaos control of Bonhoeffer-van der Pol oscilaltor using neural
networks. \textit{Chaos, Solitons and Fractals} \textbf{12}
(2001), 2395-2405.

\end{thebibliography}
\end{document}